\documentclass[
  aps,
  prl,
  reprint,
  preprintnumbers,
  showpacs,
  groupedaddress,
  amsmath,
  amssymb,
  floatfix]{revtex4-1}
\usepackage{graphicx,epsfig,dcolumn,multirow}
\usepackage{natbib}
\newcolumntype{d}[1]{D{.}{.}{#1}}

\RequirePackage{xspace}

\newcommand{\simgt}{\,\hbox{\lower0.6ex\hbox{$\sim$}\llap{\raise0.6ex\hbox{$>$}}}\,}
\newcommand{\simlt}{\,\hbox{\lower0.6ex\hbox{$\sim$}\llap{\raise0.6ex\hbox{$<$}}}\,} 

\usepackage{pstricks}
\newcmykcolor{darkgreen}{1 0 0.6 0.5}  

\begin{document}

\preprint{IPPP/20/26}
\preprint{SI-HEP-2020-014}
\preprint{SFB-257-P3H-20-031}

\title{
Renormalisation scale setting for D-mixing 
}

\author{Alexander Lenz$^{\,a, \, b}$,  Maria Laura Piscopo$^{\,a}$ and Christos Vlahos$^{\,a}$}
\vspace{0.6cm}

\affiliation{
\mbox{$^{a}$  IPPP, Department of Physics, University of Durham, DH1
              3LE, UK}
\mbox{$^{b}$  Physik Department, Universit{\"a}t Siegen, Walter-Flex-Str. 3, 57068 Siegen, Germany}
\mbox{email: alexander.josef.lenz@gmail.com, maria.l.piscopo@durham.ac.uk, christos.vlahos@durham.ac.uk}
}
\date{\today}

\begin{abstract}
A naive application of the  heavy quark expansion (HQE) yields  theory estimates for the
decay rate of neutral $D$ mesons that are four orders of magnitude below the experimental determination. It is well known that this huge suppression results from severe GIM cancellations. 
We find that this mismatch can be solved by individually choosing the renormalisation scale of 
the different internal quark contributions.
For $b$ and $c$ hadron lifetimes, as well as for the decay 
rate difference of neutral $B$ mesons the effect of our scale setting procedure lies within 
the previously quoted theory uncertainties, while we get enlarged theory uncertainties for the 
semileptonic CP asymmetries in the $B$ system. 
\end{abstract}


\maketitle

\section{Introduction}
An improvement of our theoretical understanding of charm physics is crucial to
make use of the huge amount of current and future experimental charm data obtained
by LHCb \cite{Bediaga:2018lhg}, BESIII \cite{Ablikim:2019hff} and Belle II \cite{Kou:2018nap}.
The recent discovery of direct CP violation in the charm system by the LHCb collaboration
\cite{Aaij:2019kcg} is an example of this necessity. Briefly after the announcement of a non-vanishing measurement of
$\Delta A_{CP} = A_{CP} (D^0 \to K^+ K^-) -  A_{CP} (D^0 \to \pi^+ \pi^-)$ both theory papers
arguing for a beyond standard model (BSM)
\cite{Chala:2019fdb,Dery:2019ysp} (partly based on the calculation of Ref.~\cite{Khodjamirian:2017zdu})
and a standard model (SM) 
\cite{Li:2019hho,Grossman:2019xcj,Cheng:2019ggx,Soni:2020kse}
origin of this measurement appeared (a summary of references investigating a previous claim for evidence of
CP violation can be found in Ref.~\cite{Lenz:2013pwa}).
Thus a decisive conclusion about the
potential size of the SM contribution to $\Delta A_{CP}$ is mandatory to fully exploit
the significant experimental progress in this field.
A long-standing puzzle in this regard is the theoretical description of mixing of neutral
$D$ mesons. Charm-mixing is by now experimentally well established and HFLAV \cite{Amhis:2019ckw} finds as an average 
of \cite{Aitala:1996vz,Cawlfield:2005ze,Aubert:2007aa,Bitenc:2008bk,Aitala:1996fg,Godang:1999yd,Link:2004vk,Zhang:2006dp,Aubert:2007wf,Aaltonen:2013pja,Ko:2014qvu,Aaij:2017urz,Aubert:2008zh,Aaij:2016rhq,Aitala:1999dt,Link:2000cu,Csorna:2001ww,Lees:2012qh,Aaltonen:2014efa,Ablikim:2015hih,Aaij:2015yda,Staric:2015sta,Aaij:2017idz,Aaij:2018qiw,Aubert:2007if,Canto:2013fza,Aaij:2019kcg,delAmoSanchez:2010xz,Peng:2014oda,Aaij:2015xoa,Aaij:2019jot,delAmoSanchez:2010xz,Zupanc:2009sy,TheBABAR:2016gom,Asner:2012xb}:
\begin{equation}
  x = \frac{\Delta M_D      }{  \Gamma_{D^0}} = 0.39^{+0.11}_{-0.12} \% \, \,  ,
  y = \frac{\Delta \Gamma_D }{2 \Gamma_{D^0}} = 0.651^{+0.063}_{-0.069} \% \, ,
\end{equation}
where $\Delta M_D$ is the mass difference of the two mass eigenstates of the neutral $D^0$ mesons
and $\Delta \Gamma_D$ the corresponding decay rate difference. 
However, theory predictions for $x$ and $y$ cover a vast range of values - differing by 
several orders of magnitude, see e.g. the compilations of theory predictions in Refs.~\cite{Nelson:1999fg,Petrov:2003un}. 
Future measurements will not only increase the precision of $x$ and $y$, but  also give stronger bounds or even a measurement of the CP violation in mixing \cite{Cerri:2018ypt} encoded e.g. in the phase $\phi_{12}$, which is currently constrained to be within $[ -2.5^\circ, 1.8^\circ]$
\cite{Amhis:2019ckw}. A~reliable range of potential SM values is pivotal to benefit from the coming experimental
improvements.

\section{HQE}

The heavy quark expansion (HQE)
\cite{Khoze:1983yp,Shifman:1986mx,Bigi:1992su,Bigi:1992su,Blok:1992hw,Blok:1992he,Chay:1990da,Luke:1990eg}
(see Ref.~\cite{Lenz:2014jha} for a recent overview)
describes the total decay rate of heavy hadrons and the decay rate difference of heavy neutral mesons 
as an expansion in inverse powers of the heavy quark mass.
In the case of
$B_s$-mixing and $b$-hadron lifetimes
the HQE predicts values \cite{Kirk:2017juj,Lenz:2014jha,Lenz:2019lvd,Davies:2019gnp,Dowdall:2019bea,King:2019lal,DiLuzio:2019jyq}  which are in good agreement with the experimental ones \cite{Amhis:2019ckw}: 
\begin{displaymath}
  \begin{array}{|c||c|c|}
    \hline
                                       & \rm HFLAV \, \, 2019 & \rm HQE \, \, 2019
    \\
    \hline
    \hline
    \frac{\tau(B_s)}{\tau(B_d)}        & 0.994(4)              & 1.0007(25)
      \\
    \hline   
    \frac{\tau(B^+)}{\tau(B_d)}        &1.076(4)               &1.082^{+0.022}_{-0.026}
    \\
    \hline
    \frac{\tau(\Lambda_b)}{\tau(B_d)} &  0.969(6)              & 0.935(54)
    \\
    \hline
    \Delta \Gamma_{B_s}                &0.091(13) \mbox{ps}^{-1} &0.090(5) \mbox{ps}^{-1}
    \\
    \hline
  \end{array}
  \end{displaymath}
This impressive result
when the expansion parameter is $\Lambda/m_b$ ($\Lambda$ denotes an hadronic scale of the
order of $\Lambda^{{\rm QCD}}$)
suggests that one might still get reasonably 
well-behaving estimates moving to the charm system, where the expansion parameter increases by a factor of three.
For the lifetime ratio $\tau(D^+)/ \tau(D^0)$ both NLO-QCD corrections to the dimension-six contribution
\cite{Lenz:2013aua} and 
values for the
non-perturbative
matrix elements of 
four 
quark operators \cite{Kirk:2017juj} are known - for all other charm hadrons this is not yet the case, 
thus, corresponding theory estimates have to be taken with care - and one finds indeed a nice agreement 
within the huge theory uncertainties:
\begin{eqnarray}
 {\frac{\tau(D^+)}{\tau(D^0)}}\Biggl|^{\rm HFLAV \, \, 2019}  \hspace{-1.5cm} = 2.536(19)  \, ,
 & \, \, \, \, \, \, &
 {\frac{\tau(D^+)}{\tau(D^0)}}\Biggl|^{\rm HQE \, \, 2019}   \hspace{-1.2cm} = 2.7^{+0.7}_{-0.8} \, .
\end{eqnarray}
Hence, it is quite surprising that a naive application of the HQE fails completely for $D$-mixing.
\section{Charm mixing}
Diagonalising the two dimensional mixing matrix of the $D^0$ and the $\bar{D}^0$ meson - containing the off-diagonal matrix elements $M_{12}$ and $\Gamma_{12} $ - one gets
\begin{equation}
    x_{12} = \frac{2 \, |M_{12}|}{\Gamma_{D^0}}, \, \,
    y_{12} = \frac{2 \, |\Gamma_{12}|}{\Gamma_{D^0}}, \, \,
   \phi_{12} = \arg \left( \frac{M_{12}}{\Gamma_{12}} \right) \,,
\end{equation}
while $x$ and $y$ depend on both $M_{12}$ and $\Gamma_{12}$ . 
The calculation of $M_{12}$ is beyond the scope of the present 
work hence we can only determine one contribution to
the mixing phase, also for $\Delta \Gamma_D$ we will use the bound
$\Delta \Gamma_D \leq 2 \, | \Gamma_{12}| $ 
(see
e.g. Refs.~\cite{Nierste:2009wg,Jubb:2016mvq}).
\begin{figure}
  \includegraphics[width=0.49\textwidth, angle = 0]
                  {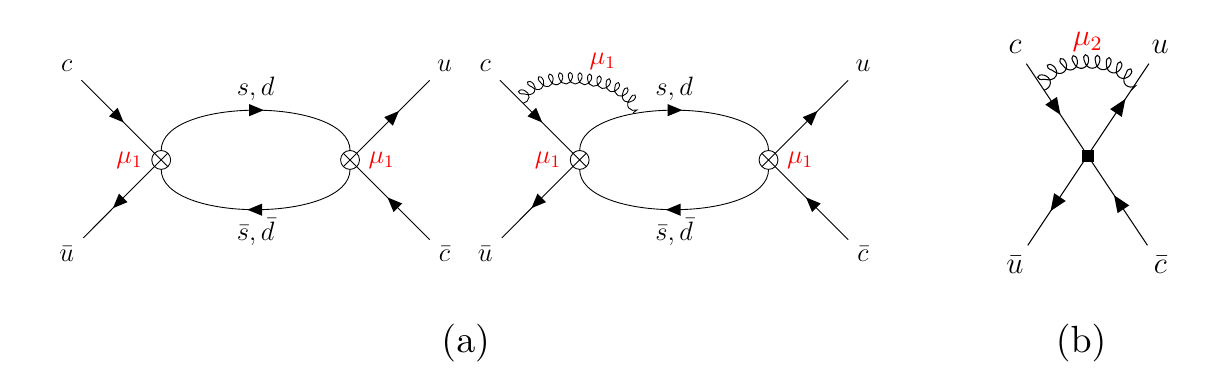}
  \caption{ (a) Diagrams describing mixing of neutral 
  $D$ mesons in the "full" theory at LO-QCD (left) and NLO-QCD (right), with intermediate $s \bar{s}$,
  $s \bar{d}$, $d \bar{s}$ and $d \bar{d}$ states. 
  The crossed circles denote the insertion of $\Delta C=1$ operators of the effective Hamiltonian. The dependence on the renormalisation scale $\mu_1$ in the Wilson coefficients cancels against the $\mu_1$~dependence of the QCD corrections. 
  (b) Diagram describing mixing of neutral $D$ mesons  at NLO-QCD in the HQE. The full dot indicates the insertion of $\Delta  C = 2$ operators. The dependence on the renormalisation scale $\mu_2$ cancels between the QCD corrections and the matrix elements.
  }
  \label{fig:mixing}
\end{figure}
Within the HQE $\Gamma_{12}$ is expanded as:
\begin{eqnarray}
  \Gamma_{12} = \left[ \Gamma_3^{(0)} + \frac{\alpha_s}{4 \pi}\,  \Gamma_3^{(1)}
    + \ldots \right] 
  \frac{\langle Q_6 \rangle}{m_c^3}
   + \ldots \, ,
\label{eq:HQE}
\end{eqnarray}
where the ellipsis stands for terms of higher order.
Eq.~(\ref{eq:HQE}) is diagrammatically represented in Fig.~\ref{fig:mixing}. The product of $\Delta C = 1$ operators in the effective Hamiltonian ("full" theory) is matched into local $\Delta C = 2$ operators in the HQE. 
The expressions for $\Gamma_3^{(i)}$ can be simply obtained from the corresponding ones for $B$-mixing given in Refs.~\cite{Beneke:1996gn,Beneke:1998sy,Dighe:2001gc,Beneke:2003az,Ciuchini:2003ww,Lenz:2006hd} while
the matrix elements of the dimension-six operators have been determined in e.g. Refs.~\cite{Kirk:2017juj,Bazavov:2017weg}.
Experiments yield a small value for the decay rate difference
$\Delta \Gamma_D^{\rm Exp} = 2 y / \tau (D^0) 
= 0.032 \pm 0.003$ ps$^{-1}$, 
which leads to
the following bound $\Delta \Gamma_D^{\rm Exp} \geq 0.028 $ ps$^{-1}$ at 1 standard deviation.
Below we will investigate the quantities
\begin{equation}
 \alpha = - \arg (\Gamma_{12}) \, , \hspace{1cm} \Omega = \frac{2 \, | \Gamma_{12}|^{\rm SM}}{  0.028 \, \mbox{ps}^{-1}} \,,
  \end{equation}
where $\alpha$ contributes to CP violation in mixing and values of 
$\Omega$ smaller than one indicate a failure of our theoretical
framework to describe $D$-mixing
within the  one sigma range. A naive application of the HQE 
leads 
to $\Omega = 3.4 \cdot 10^{-5}$ at LO-QCD 
($6.2 \cdot 10^{-5}$ at NLO-QCD), i.e. the HQE prediction of the decay rate difference is more than four orders of 
magnitude smaller than the experimental determination. 
Correspondingly the phase $\alpha$ is very large, i.e. $\alpha =  93^\circ $ at LO-QCD
($\alpha =  99^\circ $ at NLO-QCD).
By default in our numerical analysis we use PDG \cite{Tanabashi:2018oca}
values for the quark ($\overline{\mbox{MS}}$) and meson masses as well as for 
the strong coupling, CKM elements from Ref.~ \cite{Charles:2004jd},
non-perturbative matrix elements from Ref.~\cite{Kirk:2017juj} 
and the $D^0$ decay constant from Ref.~\cite{Aoki:2019cca}.
\section{GIM in D-mixing}
In order to better understand the peculiarities of D-mixing we decompose $\Gamma_{12}$
according to the flavour of the internal quark pair. The three contributions are denoted $\Gamma_{12}^{ss}$, $\Gamma_{12}^{dd}$ and $\Gamma_{12}^{sd}$:
\begin{eqnarray}
  \Gamma_{12} & = & - \left( \lambda_s^2 \, \Gamma^{ss}_{12}   + 2 \,  \lambda_s \lambda_d \, \Gamma^{sd}_{12} 
  +  \lambda_d^2 \, \Gamma^{dd}_{12}  \right)
  \nonumber
   \\
  & = &
  - \,  \lambda_s^2           \left( \Gamma^{ss}_{12} - 2   \Gamma^{sd}_{12} + \Gamma^{dd}_{12}  \right)
   \label{eq:GIM}
   \\
  & &  + \, 2 \lambda_s \lambda_b \left( \Gamma^{sd}_{12} - \Gamma^{dd}_{12}                        \right)
  -  \lambda_b^2 \Gamma^{dd}_{12}.
  \nonumber
\end{eqnarray}
$\lambda_q = V_{cq} V_{uq}^*$ is the CKM element and we have used the unitarity relation 
$\lambda_d+\lambda_s+\lambda_b=0$ to eliminate $\lambda_d$. Eq.~(\ref{eq:GIM}) shows very pronounced 
hierarchies:
  \begin{eqnarray}
  -\lambda_s^2 & = & -4.791 \cdot 10^{-2} + 3.094 \cdot 10^{-6} I,
  \\
  + 2 \lambda_s \lambda_b & = &
  +2.751  \cdot 10^{-5}  +
  6.121  \cdot 10^{-5}  I,
  \\
  -\lambda_b^2 & = & +1.560 \cdot 10^{-8} - 1.757 \cdot 10^{-8} I.
\end{eqnarray}
The CKM factor in the first term of Eq.~(\ref{eq:GIM}) has by far the largest real part, while the second term
has actually the largest imaginary part - it should thus
be important for the determination of the potential size
of CP violation in mixing.
Since the relative imaginary part of $\lambda_b$ is much
larger than that of $\lambda_s$ we suggest to keep all terms in  Eq.~(\ref{eq:GIM}). 
Furthermore extreme GIM cancellations \cite{Glashow:1970gm} affect
the coefficients of the CKM elements in Eq.~(\ref{eq:GIM}). 
Expanding in the small mass parameter 
$z =  m_s^2/  m_c^2$ we find at LO-QCD (top line)
and at NLO-QCD (lower line):
\begin{eqnarray}
  { \Gamma^{ss}_{12}} & = &  
  \left\{ 
  \begin{array}{ll}
  { 1.62 - 2.34 \, z - 5.07 \, z^2 + \ldots } 
  \, ,
  \\[1mm]
  { 1.42 - 4.30 \, z - 12.45 \, z^2  + \ldots } 
  \, ,
  \end{array}
  \right.
  \label{noGIM}
  \\
  { \Gamma^{sd}_{12}  - \Gamma^{dd}_{12}}  & = & \left\{ 
  \begin{array}{ll}
  { - 1.17 \, z - 2.53 \, z^2 + \ldots  } 
  \, , \\[1mm]
  { -2.15 \, z - 6.26 \, z^2 + \ldots }
  \, ,
  \end{array}
  \right.
  \label{littleGIM}
  \\
  {\Gamma^{ss}_{12}  - 2  \Gamma^{sd}_{12} + 
  \Gamma^{dd}_{12}}  & = & 
  \left\{ 
  \begin{array}{ll}
  { - 13.38 \, z^3 + \ldots } 
  \, , 
  \\[1mm]
  { 0.07 \, z^2 - 29.72 \, z^3 + \ldots }
  \, .
  \end{array}
  \right.
\label{crazyGIM}
\end{eqnarray}
It was observed before \cite{Golowich:2005pt,Bobrowski:2010xg} that 
QCD corrections lower the GIM suppression by one power of $z$.   
The peculiarity of Eq.~(\ref{eq:GIM}) is that
the CKM dominant factor $\lambda_s^2$ multiplies the extremely GIM suppressed term given in
Eq.~(\ref{crazyGIM}), the CKM suppressed factor $\lambda_s \lambda_b$ multiplies the GIM suppressed term 
given in Eq.~(\ref{littleGIM}) and the very CKM suppressed factor $\lambda_b^2$ multiplies $\Gamma_{12}^{dd}$, 
where no GIM suppression is present. Thus the three
contributions in  Eq.~(\ref{eq:GIM}) have actually a
similar size:
\begin{eqnarray}
\Gamma_{12} & = & \left( 2.08 \cdot10^{-7} - 1.34 \cdot 10^{-11} I \right)
\mbox{(1st term)}
\nonumber
\\
&  - &\left( 3.74 \cdot 10^{-7} + 8.31 \cdot 10^{-7} I \right)
\mbox{(2nd term)}
\nonumber
\\
&+ &\left( 2.22 \cdot 10^{-8} - 2.5 \cdot 10^{-8} I \right)
\mbox{(3rd term)}.
\end{eqnarray}
It is also clear that a sizeable phase in 
$D$-mixing can only arise, if the slightly GIM suppressed
term is enhanced. 
Different solutions have been suggested in order to explain the mismatch between the HQE prediction and experimental determination.
i) Higher orders in the HQE could be less affected by GIM suppression 
\cite{Georgi:1992as,Ohl:1992sr,Bigi:2000wn} - first estimates of the dimension nine contribution 
\cite{Bobrowski:2012jf} to $D$-mixing show indeed such an enhancement, but not on a scale to reproduce 
the experimental number. For a final conclusion about this possibility a full determination of 
dimension nine and twelve would be necessary.
ii) Large violations of quark-hadron duality are excluded by the many successful tests of the HQE as stated above. 
In Ref.~\cite{Jubb:2016mvq} it was  shown that violations as small as 20 per cent could be sufficient
to explain the experimental value of $D$ mixing.
iii) The HQE is not applicable and we have to rely on different methods, like summing over the exclusive decays channels 
contributing to the decay rate difference, see e.g. 
Refs.~\cite{Falk:2001hx, Cheng:2010rv,Jiang:2017zwr}. 
\section{Alternative Scale Setting}
\begin{figure}
    \centering
    \includegraphics[scale = 0.72 ]{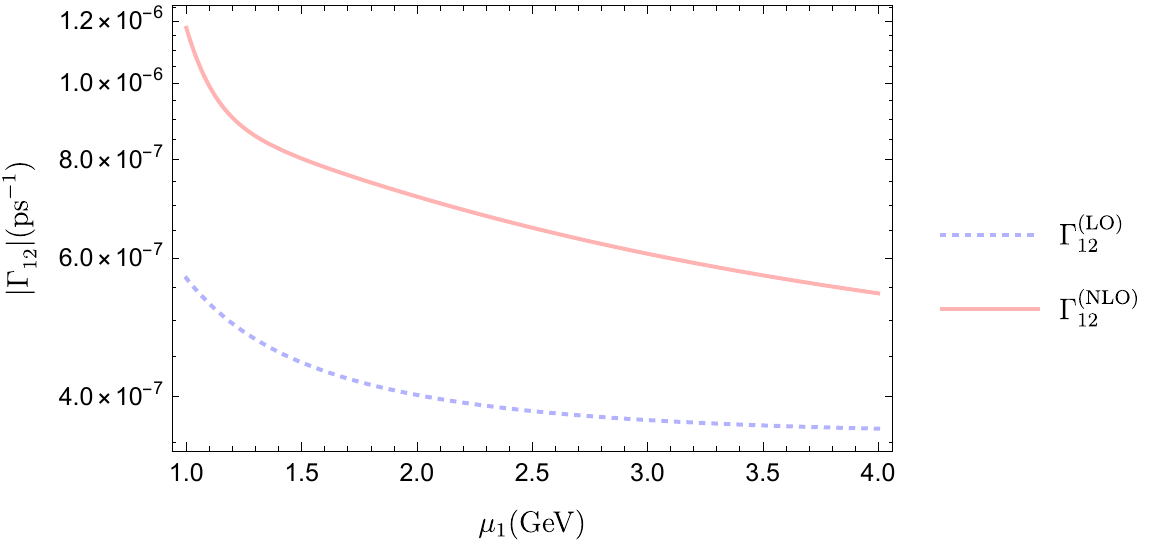}
    \caption{Comparison of $\mu_1$-dependence of $|\Gamma_{12}|$ at LO-QCD (dotted blue) and NLO-QCD (solid pink).}
    \label{fig:Omega-lo-nlo}
\end{figure}

In $\Gamma_{12}$ the two renormalisation scales
$\mu_1$ and $\mu_2$
are arising, see Fig.~\ref{fig:mixing}.
The dependence on $\mu_1$ in the $\Delta C =~1$ Wilson
coefficients of the effective Hamiltonian
cancels, up to terms of higher order, the corresponding
dependence of the radiative corrections to the diagrams, Fig.~\ref{fig:mixing}.(a).
Similarly the dependence on $\mu_2$ arises from
loop-corrections to the HQE diagrams,
Fig.~\ref{fig:mixing}.(b) and cancels the corresponding
dependence of the matrix elements of the $\Delta C=2$
four quark operators. We will 
not discuss the $\mu_2$-dependence any further since this cancellation is very effective. For the $\mu_1$-dependence,
in the $B_s$ system the cancellation is
numerically only weakly realised when 
moving from LO-QCD to NLO-QCD, see Refs.~\cite{Asatrian:2017qaz,Asatrian:2020zxa}.
This indicates the importance of higher order 
corrections and first steps in that direction show 
indeed large NNLO-QCD effects
\cite{Asatrian:2017qaz,Asatrian:2020zxa}. 
In the $D$ system a reduction of the $\mu_1$-dependence, when
moving from LO-QCD to
NLO-QCD, is present in the individual
contributions $\Gamma^{ss,sd,dd}_{12}$ but not in 
$\Gamma_{12}$, see Fig.~\ref{fig:Omega-lo-nlo}, which seems to be again a consequence of the 
severe GIM cancellations.
Making the scale dependence explicit we can write:
\begin{eqnarray}
\! \! \Gamma_{12} & = & \! \! \! \! \! \! \! \! \! \! \!
\sum \limits_{q_1q_2 =ss,sd,dd} \! \! \! \! \! \! \! \!
\Gamma_3^{q_1q_2} (\mu_1^{q_1q_2},\mu_2^{q_1q_2}) \langle Q \rangle (\mu_2^{q_1q_2})
\frac{1}{m_c^3} + \ldots \, 
\end{eqnarray}
In general different internal quark pairs contribute to
different decay channels of the $D^0\, (\bar D^0)$ meson
e.g. $s \bar{s}$ 
to a $K^+ K^-$  
final state and $s \bar{d}$ to a $\pi^+ K^-$ 
final state.
For each of these different observables the choice of the renormalisation scales is a priori arbitrary,
nevertheless one typically fixes
$\mu_x^{ss} = \mu_x^{sd} = \mu_x^{dd} = \mu$ 
which is then 
chosen to be equal to the mass of the decaying heavy 
quark, i.e. $\mu = m_Q$ for $Q$ quark decays,
to minimize terms of the form 
$\alpha_s(\mu) \ln (\mu^2/m_Q^2)$. 
Uncertainties due to unknown 
higher order corrections are estimated varying $\mu$
between $m_Q/2$ and $2\, m_Q$ - in the case of the 
charm quark 
we fix the lower bound to $1$ GeV 
in order to still ensure 
reliable perturbative results.
\\
Here we propose two different 
ways to
treat the renormalisation 
scale $\mu_1^{q_1 q_2}$, both will reduce
the mismatch between the HQE prediction and the experimental determination of $D$-mixing,
while leaving the other HQE predictions unchanged:
i) $\mu_1^{ss}$,  
$\mu_1^{sd}$ and $ \mu_1^{dd}$ are set to the common scale $m_c$ but varied independently 
between $1$~GeV~and $2 m_c$.  
ii) $\mu_1^{ss}$,  
$\mu_1^{sd}$ and $ \mu_1^{dd}$ are set to different scales
according to the
size of the available phase
space. 
In particular we will evaluate $\Gamma^{ss}_{3}$  at the scale
$\mu_1^{ss} = \mu - 2 \epsilon$,
$\Gamma^{sd}_{3}$~at the scale
$\mu_1^{sd} = \mu -  \epsilon$ and $\Gamma^{dd}_{3}$ at 
the scale $\mu_1^{dd} = \mu$, where $\epsilon$ is 
related to the kinematics of the decays.
If $\epsilon$ is not too large, then both methods will yield results for the individual
$\Gamma^{ss}_{3}$, $\Gamma^{sd}_{3}$ and  $\Gamma^{dd}_{3}$ which lie within the usually 
quoted theory uncertainties obtained
following the prescription stated above,
but they
will clearly affect in a sizeable way the severe GIM cancellations in Eqs.~(\ref{littleGIM}) and (\ref{crazyGIM}).
The first method gives
a considerably enhanced range 
of values for $\Omega$:
\begin{equation}
\Omega \in [4.6 \cdot 10^{-5} , 1.3] \, ,
\label{sol1}
\end{equation}
which nicely covers also the experimental determination 
of the decay rate difference. 
Scanning independently over $\mu_1^{ss}$, 
$\mu_1^{sd}$ and $\mu_1^{dd}$ in 11 equidistant
steps we find that out of the 1331 points only 14 give a value of 
$\Omega < 0.001$, while 984 give a value of 
$\Omega > 0.1$.
The very small HQE
prediction seems thus to be
an artefact of fixing the scales $\mu_1^{ss}$,  $\mu_1^{sd}$ and 
$\mu_1^{dd}$ to be the same.
The range of values shown in Eq.~(\ref{sol1}) is similar even if we use the pole scheme for the quark masses, lattice results instead of the HQET results or a different $\Delta C=2$ operator basis. In all these cases the value $\Omega \geq 1$ can be obtained.
For $\alpha$
we get in general results ranging from $-\pi$ to $\pi$.
A closer look however, shows
that for $\Omega >0.5$
only values of $\alpha < 0.1^\circ $ are allowed, while 
large values of $\alpha$ can only be obtained if the 
theory prediction for $y$ is
inconsistent 
with the 
experimental determination.
The second method for the scale setting
requires the introduction of
a mass scale $\epsilon$. 
A possible estimate for the size of this parameter
could be the
strange quark mass $\epsilon = m_s \approx 0.1$~GeV or 
the phase space difference of the
corresponding exclusive
decays channels:
comparing the energy release of $D^0 \to K^+ K^-$, $M_{D^0} - 2 M_{K^+} = 0.88$ GeV, with that of $D^0 \to \pi^+ \pi^-$, $M_{D^0} - 2 M_{\pi^+} = 1.59$ GeV we might 
expect that $\epsilon \approx 0.35$~GeV.
\begin{figure}
    \centering
    \includegraphics[scale=.6]{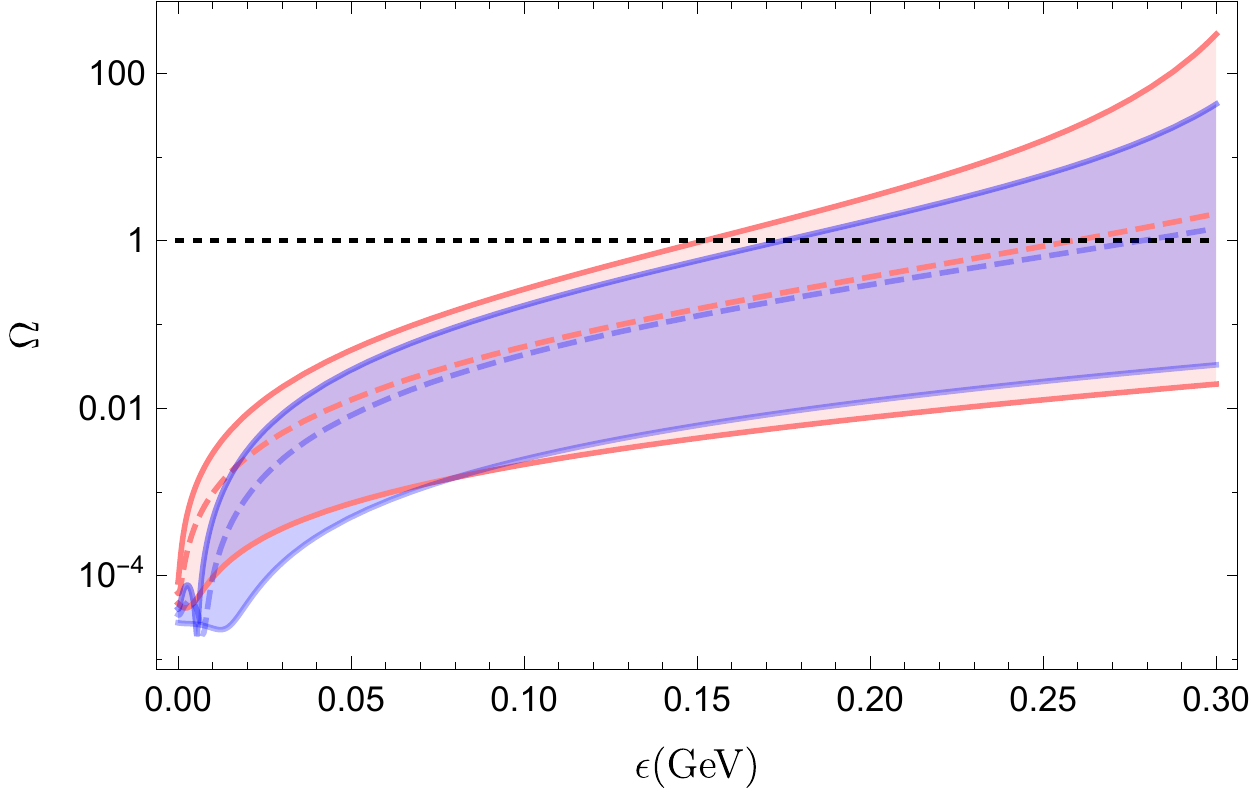}
    \caption{Comparison of the $\epsilon$ dependence of $\Omega$ at LO-QCD (blue) and  NLO-QCD (pink) for different values of $\mu$: the dashed line corresponds to $\mu =~m_c$ while the two solid lines to $\mu = 1$~GeV and $\mu = 2 m_c$.}
    \label{fig:Omega-epsilon-variation}
\end{figure}
Fig.~\ref{fig:Omega-epsilon-variation} shows how the 
HQE prediction of $\Omega$ would be affected in this
scenario. Again an enhancement up to the experimental
value is possible for values of $\epsilon \approx 0.2$~GeV.   
Finally we have to test the effect of our alternative scale setting procedure on all the other 
HQE predictions.
For the lifetimes  
(i.e.   $\tau (D^+)/ \tau(D^0)$ as well as $b$ hadron
lifetimes) and the decay rate difference 
$\Delta \Gamma_s$ no GIM-like cancellations arise and we can only get a shift within the usually quoted 
theory range.
But the semi-leptonic CP
asymmetries are governed 
by the weakly GIM suppressed contribution in $B_s$-mixing. Within the SM we get
\begin{eqnarray}
{\rm Re} \left( 
\frac{\Gamma_{12}^q}{M_{12}^q}\right)^{\rm SM} 
& = & 
- \frac{\Delta \Gamma_q}{\Delta M_q} = 
\left\{
\begin{array}{ll}
- (49.9 \pm 6.7) \cdot 10^{-4} & q = s 
\\
- (49.7 \pm 6.8) \cdot 10^{-4} & q = d 
\end{array}
\right.
\,  ,
\nonumber
\\
{\rm Im} \left( \frac{\Gamma_{12}^q}{M_{12}^q} 
\right)^{\rm SM} 
& = &  
a_{sl}^q = 
\left\{
\begin{array}{ll}
(+2.2 \pm 0.2) \cdot 10^{-5} & q = s
\\
(-5.0 \pm 0.4) \cdot 10^{-4} & q = d
\end{array}
\right.
\,  .
\end{eqnarray}
Performing the $\epsilon$ analysis
we find:
\begin{displaymath}
\begin{array}{|c||c|c|}
\hline
\epsilon \,( {\rm GeV} ) & \Gamma_{12}^s / M_{12}^s & \Gamma_{12}^d / M_{12}^d
\\
\hline \hline
0.    & {\color{blue} -0.00499 + 0.000022 I}  & {\color{blue} -0.00497 - 0.00050 I}
\\ \hline
0.2.  & {\color{blue} -0.00494 + 0.000023 I}& {\color{blue}-0.00492 - 0.00053 I}
\\ \hline
0.5.  &  {\color{blue}-0.00484} + 0.000026 I&  {\color{blue}-0.00482} - 0.00059 I
\\ \hline
1.0   &  {\color{blue}-0.00447 }+ 0.000037 I&  {\color{blue}-0.00448} - 0.00084 I
\\ \hline
1.5.  & -0.00287 + 0.000091 I&                -0.00309 - 0.0021 I
\\ \hline
\end{array}
\end{displaymath}
We see, that for $\epsilon$ values of up to 1 GeV the
predictions for the real part lie with the usually quoted
theory uncertainties (indicated in blue). 
The predictions for the semi-leptonic asymmetries can, however, be increased by almost $100 \%$ compared to the usually quoted values.
\section{Conclusions}
Our main finding is that the range of the HQE uncertainty
for $y$ is much larger than previously thought and it
covers the experimental value if we modify the usually
adopted scale setting. 
For a full solution of the $D$-mixing puzzle we
nevertheless suggest a more precise estimate of higher
order corrections in the HQE, as well as a completion of
the NNLO-QCD corrections to the leading term.
\\
In our alternative scale setting procedure we find that  a small contribution to CPV in mixing stemming from 
the decay rate can be up to one per mille within in the 
SM, which agrees with estimates made in Refs.~\cite{Kagan:2020vri,Li:2020xrz}. For a prediction of 
CP violation in mixing in addition the  contribution
coming from $M_{12}$ has to be determined. 
This might be done in future via the help of dispersion 
relations, see e.g.  Refs.~\cite{Falk:2004wg, Cheng:2010rv,Li:2020xrz}.
We would like to note that our suggested procedure
is still respecting the GIM mechanism, because for
vanishing internal strange quark mass, also the parameter
$\epsilon$ will be zero.
\\
Finally this scale setting does not affect quantities  like $\tau (D^+)/ \tau(D^0)$, $b$ hadron lifetimes and  $\Delta \Gamma_s$ outside the range of their quoted theoretical errors, but it affects the semi-leptonic CP asymmetries and we get enhanced SM ranges:
\begin{eqnarray}
a_{sl}^d  \in  [-9.2;-4.6] \cdot 10^{-4}  , \, \,
&&
a_{sl}^s  \in  [2.0; 4.0] \cdot 10^{-5} \, .
\end{eqnarray}

\section{Acknowledgements}

We thank Vladimir Braun, Marco Gersabeck, Thomas Rauh, Alexey Petrov and  Aleksey Rusov 
for helpful discussions.

\bibliography{dmixing}

\end{document}